\documentclass[twocolumn,pra]{revtex4}

\usepackage{graphicx}
\usepackage{dcolumn}
\usepackage{bm}

\begin{document}


\title{Quantum phase diffusions of a spinor condensate}
\author{S. Yi, \"O. E. M\"ustecapl{\i}o\u{g}lu, and L. You}
\affiliation{School of Physics, Georgia Institute of
Technology, Atlanta, GA 30332-0430, USA}
\date{\today}
\begin{abstract}
We discuss the quantum phases and their diffusion in a
spinor-1 atomic Bose-Einstein condensate. For ferromagnetic
interactions, we obtain the exact ground state distribution
of the phase fluctuations corresponding to the total
atom number ($N$), the magnetization (${\cal M}$),
and the alignment (or hypercharge) ($Y$) of the system.
The mean field ground state
is shown to be stable against these fluctuations,
which dynamically recover the two continuous symmetries
associated with the conservation of $N$ and ${\cal M}$
as in current experiments.

\end{abstract}

\pacs{03.75.-b, 67.90.+z} \maketitle

Since the observation of Bose-Einstein condensation of
trapped atomic clouds \cite{bec,mit}, the coherence properties
of the condensate has become the focus of many theoretical studies
\cite{lewenstein,wright,imamoglu,villain97,law,villain99,juha}.
Within the mean-field theory, it is commonly assumed that
the condensate can be described by a $U(1)$ symmetry breaking
field \cite{crispen,castin}, equivalent to a coherent state
assumption of the ground state. (see Refs. \cite{crispen,castin}
for discussions of $U(1)$-symmetric approaches).
Although quite successful in providing
theoretical understanding to many experimental
observations, such a coherent state assumption is not
necessarily consistent with real experimental situations,
where the fluctuations of the atom numbers are
difficult to control \cite{nf,nfs}.
A coherent state leads to a Poisson distribution
of atoms. For a ground state with average of $N$ atoms,
the associated number fluctuations are of the order $\sqrt{N}$.
As was
initially pointed out in Refs. \cite{lewenstein,wright}, this
number fluctuation of a coherent
state condensate leads to the
``diffusion" (or spreading) of its initial phase.
In a scalar condensate, this diffusion, a dynamic attempt to
restore the $U(1)$ symmetry of the interacting atomic system,
can be studied in terms of a zero mode, or the
Goldstone mode of the condensate \cite{lewenstein}.
More physically meaningful
discussions in terms of the relative phase of two condensates
were studied soon afterwards
\cite{imamoglu,villain97,villain99,law}. Experimentally, starting
with the remarkable direct observation of the first order
coherence in an interference experiment \cite{mitint}, direct
correlations between number and phase fluctuations
were observed with a condensate in a periodic
potential \cite{kasevich}, and more recently, in the
remarkable Mott insulating state \cite{hansch} obtained
by loading a superfluid condensate into an optical lattice \cite{zoller}.

The emergence of spinor-1 condensates \cite{kurn,mike}
(of atoms with hyperfine quantum number $F=1$) has created
new opportunities to understand quantum coherence and the
associated number/phase dynamics in a three component condensate
\cite{burnett,ashab}.
In this paper, we investigate the quantum phase dynamics of a
spinor-1 condensate due to atom number fluctuations.
We will focus on ferromagnetic interactions, when
the condensate wave functions for the three spin components
share the same spatial mode \cite{su};
in this case, the phase fluctuations translate into
fluctuations of the direction of the macroscopic condensate spin.

We consider a system of $N$ spin-1 bosonic atoms
interacting via only s-wave scattering \cite{ho,ohmi,goldstein,pu}.
A weak magnetic field $B$ (as always exists in an experiment)
fixes the quantization z-axis such that quadratic Zeeman
effect can be neglected.
In the second-quantized form, the Hamiltonian
is \cite{ho,ohmi}
\begin{eqnarray}
H&=&\sum_i\int d\,\vec r\psi_i^\dag(\vec
r)\left[-\frac{\hbar^2\nabla^2}{2M}
+V_{\rm ext}(\vec r)-\hbar\omega_L{\mathbf F}_z\right]\psi_i(\vec r)\nonumber\\
&&+\frac{c_0}{2}\sum_{i,j}\int d\,\vec r\psi_i^\dag(\vec
r)\psi_j^\dag(\vec r)
\psi_i(\vec r)\psi_j(\vec r)\nonumber\\
&&+\frac{c_2}{2}\sum_{i,j,k,l}\int d\,\vec r\psi_i^\dag(\vec r)
\psi_j^\dag(\vec r)\vec{\mathbf F}_{ik} \cdot\vec{\mathbf
F}_{jl}\psi_l(\vec r)\psi_k(\vec r),\hskip 18pt \label{ham}
\end{eqnarray}
where $\psi_j(\vec r)$ ($j=+,0,-$) denotes the annihilation
operator for the $j$-th component of a spinor-1 field. The
trapping potential $V_{\rm ext}(\vec r)$ is assumed
harmonic and spin-independent. The Larmor precessing
frequency is $\omega_L=B\mu_B/\hbar$ with $\mu_B$ the magnetic
dipole moment for state $|F=1,M_F=1\rangle$.
The pseudo potential coefficients are
$c_0=4\pi\hbar^2(a_0+2a_2)/3M$ and $c_2=4\pi\hbar^2(a_2-a_0)/3M$,
with $a_0$ ($a_2$) the s-wave scattering length for two spin-1
atoms in the combined symmetric channel of total spin $0$ ($2$).
$M$ is the mass of atom, and $\vec\mathbf F$ is the spin 1
operator \cite{pu}. Hamiltonian (\ref{ham})
is invariant under U(1) gauge transformation
$e^{i\theta}$ and SO(3) spin rotations ${\cal
U}(\alpha,\beta,\tau)=e^{-iF_z\alpha}e^{-iF_y\beta}e^{-iF_z\tau}$
(for $B=0$) \cite{ho}.
A non-zero $B$ or the conservation of magnetization
${\cal M}=N_+-N_-$ reduces the SO(3) to its subgroup SO(2)
generated by $e^{-iF_z\alpha}$.

Following the Bogoluibov theory
we assume there exist three
`large' condensate components $\phi_j$ around
which we study the small quantum fluctuations
(off-condensate excitations) via
\begin{eqnarray}
\psi_j(\vec r)=\sqrt{N}\phi_j(\vec r)+\delta\psi_j(\vec r),
\label{bog}
\end{eqnarray}
where $N\int d\vec r|\phi_j|^2=N_j$
is the number of condensed atoms in component $j$.
At near zero temperatures, higher than quadratic
terms of $\delta\psi_j(\vec r)$ in the Hamiltonian (\ref{ham})
are neglected. The assumed mean field ground state
$\sqrt{N}\phi_j(\vec r)$ breaks both continuous symmetries
U(1) and SO(2), thus we expect to observe multiple
zero energy Goldstone modes \cite{lewenstein}.

In the first order, we obtain the usual coupled
Gross-Pitaevskii equation (GPE)
for the condensate modes $\phi_j(\vec r)$.
The quantum fluctuations $\delta\psi_j$ obey the
usual Bogoluibov-de Gennes equations (BdGEs) \cite{bog}.
The condensate number fluctuations, can be studied
through the number fluctuation operators \cite{lewenstein}
\begin{eqnarray}
P_j &\equiv & \int d\,\vec r\left[\phi_j^\ast(\vec r)\delta\psi_j(\vec
r)+\phi_j(\vec r)\delta\psi_j^\dag(\vec r)\right]\nonumber\\
&\approx &{1\over \sqrt{N}}\int d\vec r [\psi_j^\dag(\vec r)
\psi_j(\vec r)-N\phi_j^*(\vec r)\phi_j(\vec r)].
\end{eqnarray}
Using the GPEs and the BdGEs \cite{bog}
their equations of motion are \cite{lewenstein},
\begin{eqnarray}
i\hbar\dot{P}_\pm
=-{i\hbar}\dot{P}_0/2,
\end{eqnarray}
from which we find that both $P_{\rm tot}=P_++P_0+P_-$ and $P_+-P_-$
are constants of motion, an obvious outcome since
the Hamiltonian (\ref{ham}) commutes with operators
of both the total number of atoms and magnetization.

We define the phase operator
\begin{eqnarray}
Q_j=i\hbar\int d\,\vec r\left[\theta_j^\ast(\vec
r)\delta\psi_j(\vec r)-\theta_j(\vec r)\delta\psi_j^\dag(\vec
r)\right],
\end{eqnarray}
with $\theta_j(\vec r)$ the associated phase mode functions.
Denote
$Q_{\rm tot}=Q_++Q_0+Q_-$,
the canonical quantization condition $[Q_{\rm tot},P_{\rm
tot}]=i\hbar$ is satisfied if the constraint $J_++J_0+J_-\equiv 1$
is enforced, where $J_j\equiv\int d\,\vec r\left[\theta_j^\ast(\vec
r)\phi_j(\vec r) +c.c.\right]$.
As for a scalar or binary condensate \cite{villain97}
\begin{eqnarray}
{dQ_{\rm tot}\over dt}=N\tilde{u}P_{\rm tot}.
\end{eqnarray}

We note that $[Q_j,P_k]=i\hbar\delta_{jk}J_j$, $[\delta\psi_j(\vec
r),P_k]=\delta_{jk}\phi_j(\vec r)$, and $[\delta\psi_j(\vec
r),Q_k]=-i\hbar\delta_{jk}\theta_j(\vec r)$. These lead to
$\delta\psi_j(\vec r)\approx \theta_j(\vec r)P_j/J_j+\phi_j(\vec
r)Q_j/i\hbar J_j$ \cite{villain97,villain99}, which
helps to define the complete dynamic
equations for the number and phase fluctuations \cite{yiL}.

Including both the initial and time phases,
the ground state wave functions can be generally expressed as
$|\phi_j(\vec r)| e^{-i(\mu-j\hbar\omega_L)t/\hbar+i\alpha_j}$
with a common chemical potential $\mu$, a Lagrange multiplier
enforcing the conservation of atom numbers.
Minimization of the total energy Eq. (\ref{ham})
leads to $\alpha_++\alpha_--2\alpha_0 =0$ \cite{isoshima}.
It was further proven in Ref. \cite{su}
that the steady state solution takes the same
spatial mode for each of its three spin components
(for ferromagnetic interactions), i.e.
\begin{eqnarray}
\phi_j(\vec r)=\sqrt{n_j}\,\phi(\vec r)e^{i\alpha_j}, \label{wave}
\end{eqnarray}
where the real-valued mode function $\phi(\vec r)$ is normalized to
unity, and governed by an equivalent scalar GPE
\begin{eqnarray}
\left[-{\hbar^2\nabla^2\over 2M}
+V_{\rm ext}(\vec r)+c_+N\phi^2(\vec r)\right]\phi(\vec r) =\mu\phi(\vec
r), \label{gpe1}
\end{eqnarray}
of a scattering length $a_2$ (note $c_+=c_0+c_2\propto a_2$).
Define $n_j=N_j/N$ and $m={\cal M}/N$ as
the relative atom numbers and relative magnetization,
$n_\pm=(1\pm m)^2/4$ and $n_0=(1-m^2)/2$ in
the ferromagnetic ground state \cite{pu,su}.

Similarly, the phase functions share the same spatial
mode and can be generally expressed as
\begin{eqnarray}
\theta_j(\vec r,t)=\sqrt{n_j}\,\theta(\vec r)
e^{-i(\mu-j\hbar\omega_L) t/\hbar+i\alpha_j},
\end{eqnarray}
with $\theta(\vec r)$ governed by the following equation
\begin{eqnarray}
\left[-{\hbar^2\nabla^2\over 2M}
+V_{\rm ext}-\mu+3c_+N\phi^2\right]\theta(\vec r)
=N\tilde{u}\phi(\vec r).
\label{eqphase}
\end{eqnarray}
The normalization $\int d\,\vec r\phi(\vec r)\theta(\vec r)=1/2$
determines the Goldstone parameter $\tilde{u}$,
which in the Thomas-Fermi (TF) limit is
$\tilde u=[c_+/(4\pi a_{\rm ho}^3/3)](a_{\rm ho}/15 a_2)^{3/5}N^{-3/5}$,
with $a_{\rm ho}$ the ground state size
$(\hbar/M\omega_{\rm ho})^{1/2}$
of a harmonic trap with a frequency
$\omega_{\rm ho}=(\omega_x\omega_y\omega_z)^{1/3}$.

Equations (\ref{gpe1}) and (\ref{eqphase}) are identical to
their counterparts for a scalar condensate \cite{lewenstein}.
In the ferromagnetic ground state,
individual atomic spins align along the same direction,
thus they collide only in the symmetric total spin $F=2$ channel.
To conserve the magnetization with respect to the $B$-field
direction, the condensate spin direction is simply tilted
at an angle $\theta=\cos^{-1}m$. If this direction is
taken as the quantization axis, then the spinor
condensate behaves essentially as a scalar one \cite{ho,su,ueda}
when the $B$-field is zero.


With the mode functions for atom number and
phase fluctuations, we find the zero mode dynamics
can be expressed in terms of an
associated Hamiltonian. After laborious calculation
we arrive at \cite{yiL}
\begin{eqnarray}
\frac{H_{\rm zero}}{N}
&=&{\mathbf p}'^T{\mathbf{\cal A}{\mathbf p}'} +{\mathbf
q}'^T{\mathbf{\cal B}{\mathbf q}'},\label{hzero2}
\end{eqnarray}
with ${\mathbf p}'^T=(p_+',p_0',p_-')$, ${\mathbf
q}'^T=(q_+',q_0',q_-')$, redefined as canonically
conjugated variables $p_j'\equiv P_j$ and $q_j'\equiv
Q_j/J_j$.
${\mathbf{\cal A}}$ \& ${\mathbf{\cal
B}}$ are two Hermitian and positive-definite
matrices involving only parameters of the system.
Thus $H_{\rm zero}\ge 0$, the assumed ground state,
a mean field symmetry breaking state with coherent
condensate amplitudes
$\sqrt{N}\phi_j(\vec r)$ is stable.
The associated quantum fluctuations of atom numbers
can be studied with the linearization approximation Eq. (\ref{bog}).
Matrices ${\mathbf{\cal A}}$ \& ${\mathbf{\cal
B}}$ are simplified with
 ${\cal O}_{\phi\theta}\equiv \int d\,\vec
r\phi^2(\vec r)\theta^2(\vec r)$ and ${\cal
O}_{\phi\phi}\equiv\int d\,\vec r\phi^4(\vec r)$.
We also find
$\tilde{u}=\varepsilon+4c_+{\cal O}_{\phi\theta}$
with
$N\varepsilon={2}\int d\vec r\theta(\vec r)\left[{\cal L}
-\mu+c_+N\phi^2(\vec r)\right]\theta(\vec r)$ a
non-negative quantity.

\begin{figure}
\centering
\includegraphics[width=3.25in]{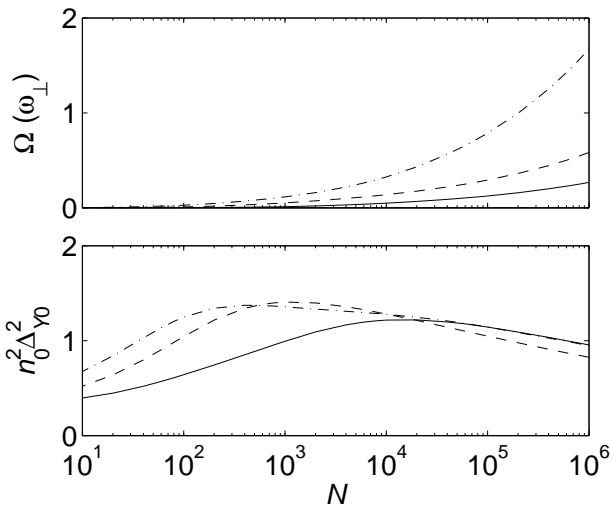}
\caption{ The N-dependence of $\Omega$ and $3n_0^2\Delta_{Y0}^2$
for a $^{87}$Rb condensate in a cylindrically symmetric
trap with $\omega_x=\omega_y=\omega_\perp=(2\pi) 100$ (Hz) and
$\omega_z=\lambda\omega_\perp$.
$\lambda=0.1$ (solid line),
$\lambda=1.0$ (dashed line),  and $\lambda=10.0$ (dash-dotted line).
}
\label{fig2}
\end{figure}

We note that matrix $\mathbf{\cal B}$ is diagonalized by
an orthogonal transformation
$q_N=(q'_++q'_0+q'_-)/\sqrt{3}$,
$q_M=(q'_+-q'_-)/\sqrt{2}$, and
$q_Y=(q'_+-2q'_0+q'_-)/\sqrt{6}$
 \cite{ozgur}. The corresponding $p_N$, $p_M$, and $p_Y$
are respectively fluctuations of the total
number of atoms, the magnetization, and the alignment.
With these new collective operators, the zero mode Hamiltonian becomes
\begin{eqnarray}
\frac{H_{\rm zero}}{N}&=&ap_N^2+bp_M^2+cp_Y^2\nonumber\\
&&+\alpha p_Np_Y+\beta p_Mp_Y
+\gamma p_Np_M +\eta q_Y^2,\label{hzero}
\end{eqnarray}
where all the coefficients are listed in the Appendix.
Replacing $p_j$ by $-i\hbar\partial/\partial q_j$,
Hamiltonian (\ref{hzero}) leads to the ground state
distribution of fluctuations
\begin{eqnarray}
\varphi_0(q_N,q_M,q_Y)={1\over \sqrt{\sqrt{\pi}\,\hbar\Delta_{Y0}}}
\exp\left(-\frac{q_Y^2}{2\hbar^2\Delta_{Y0}^2}\right),\hskip 12pt
\end{eqnarray}
with a width
$$\Delta_{Y0}^2
=\frac{1}{3n_0^2} \left(\frac{\varepsilon-4c_2{\cal O}_{\phi\theta}}
{-c_2{\cal O}_{\phi\phi}}\right)^{1/2},$$
and the zero point energy
$E_0\equiv \hbar\Omega/2=
N[-c_2{\cal O}_{\phi\phi}(\varepsilon-4c_2{\cal O}_{\phi\theta})]^{1/2}$.
We find $n_0^2\Delta_{Y0}^2=\sqrt{7/10}/3$ and
$\Omega\propto -({c_2/\pi a_{\rm ho}^3})
({a_{\rm ho}/a_2})^{3/5}N^{2/5}$
in the TF limit (and also taking the small $\varepsilon\to 0$).
Figure \ref{fig2} shows selected results for the
$N$-dependence of $\Omega$ and $\Delta_{Y0}^2$.
While $\Omega$ increases monotonically with $N$,
the ground state
width $\Delta_{Y0}^2$ for $q_Y$ peaks
at some intermediate values of atom numbers and
eventually saturates in the limit of large $N$
(or strong interactions).

The above ground state distribution
of $H_{\rm zero}$ can easily be understood. For
a condensate with fixed total number of atoms and
magnetization, the conservations of $N$ and ${\cal M}$
require $\langle p_N^2\rangle= \langle
p_M^2\rangle=0$, which lead to $\langle q_N^2\rangle= \langle
q_M^2\rangle=\infty$, i.e.,
completely diffused phases; The
ferromagnetic interaction, nevertheless, prepares a correlated
ground state such that both distributions for $p_Y$ and $q_Y$
take a Gaussian form
with $\langle p_Y^2\rangle= 1/2\Delta_{Y0}^2$ and
$\langle q_Y^2\rangle=\hbar^2\Delta_{Y0}^2/2$. Such
a distribution in $p_Y$ and $q_Y$ is in fact
the minimal uncertainty coherent state consistent
with the symmetry breaking ground state.
However, it is experimentally difficult to produce
such a condensate having fixed atom numbers and
phase fluctuations. For any initial state,
the dynamic solution
${\mathbf x}(t)={\mathbf{\cal T}}(t){\mathbf x}(0)$
as governed by the Hamiltonian Eq. (\ref{hzero}) for
${\mathbf
x}^T(t)=[p_N(t),p_M(t),p_Y(t),q_N(t),q_M(t),q_Y(t)]$
can be used. This solution has a
simple structure \cite{yiL};
in addition to oscillating terms of the forms
$\cos\Omega t$ and $\sin\Omega t$, the phase fluctuations of $q_N$
and $q_M$ also contain diffusion terms proportional to $Nt$,
which indicates the linearization approximation
of Eq. (\ref{bog}) is valid only for a finite duration.

\begin{figure}
\centering
\includegraphics[width=3.25in]{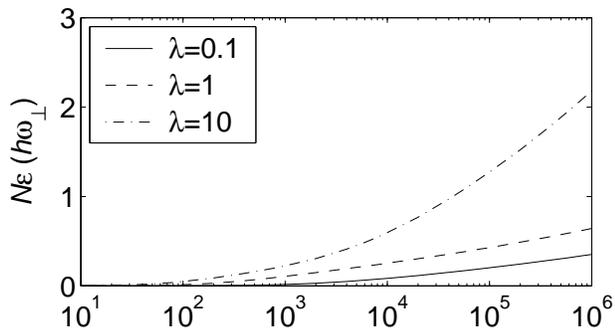}
\caption{The same as in Fig. \ref{fig2} but for
$N\varepsilon$.} \label{fig3}
\end{figure}

Finally, let's consider the diffusion of the direction of the
condensate spin $\vec f(\vec r,t)\equiv\sum_{ij}\psi_i^\dag(\vec r,t)
\vec{\mathbf F}_{ij}\psi_j(\vec r,t)$. Because all three
spin components share the same spatial mode function \cite{su},
individual spins of Bose condensed atoms are parallel
for ferromagnetic interactions,
i.e. they act as a macroscopic magnetic dipole pointing
along the same direction (independent of the spatial coordinates):
$\vec f_0(\vec r,t)
\propto (\sqrt{1-m^2}\cos\Theta,
-\sqrt{1-m^2}\sin\Theta, m)$
while undergoing precessing with respect to the z-axis.
$\Theta=\alpha_+-\alpha_0+\omega_Lt$.
As the phase dynamics attempts to restore the U(1)
and SO(2) symmetries of the system, their respective
initial values become irrelevant.
The fluctuation becomes
$\delta\vec f(t)=\int d\vec r\,[\vec f(\vec r,t)-\vec f_0(\vec r,t)]$
with a zero average.
Using $\sigma_N^2$, $\sigma_M^2$, and $\sigma_Y^2$
to denote the initial variances of $N$, ${\cal M}$,
and $Y$, we find in spherical
coordinates $(\hat r,\hat\theta,\hat\phi)$
(note $\sin\theta=\sqrt{1-m^2}$), $\delta f_r(t)\equiv\delta
f_r(0)$ and $\delta f_\theta(t)\equiv\delta f_\theta(0)$.
Both are fixed constants
\begin{eqnarray}
\left\langle\left[\delta f_r(t)\right]^2\right\rangle&=&
3\sigma_N^2,\\
\left\langle\left[\delta
f_\theta(t)\right]^2\right\rangle&=&\frac{1}{1-m^2}
\left( 3m^2\sigma_N^2+2\sigma_M^2\right),
\end{eqnarray}
(due to conservations of $N$ and ${\cal M}$)
when respective fluctuations in $N$, ${\cal M}$,
and $Y$ are uncorrelated. In the azimuthal $\hat\phi$ direction
a simple phase diffusion results
\begin{eqnarray}
\delta f_\phi(t)=\delta f_\phi(0)-R_d t,
\end{eqnarray}
with
$\delta f_\phi(0)={\sqrt{1-m^2}}\left[\sqrt{2}q_M(0)
+\sqrt{6}mq_Y(0)\right]/2\hbar,$
and a diffusion rate
$R_d={2\varepsilon N}\delta f_\theta(0)/\hbar$,
proportional to $\delta f_\theta(0)$, (which can be
explained in terms of the single axis twisting of
$T_z^2$ in the isospin subspace of a spior-1 condensate \cite{ozgur}).
The $N$- and $\lambda$-dependence
of the diffusion parameter $N\varepsilon$ are
shown in Fig. \ref{fig3} for a $^{87}$Rb condensate.
In the TF limit and applying the results of Ref. \cite{dalfovo},
we find that $\varepsilon=({4\hbar^2}/{5MR^2})
\ln({R}/{1.3a_{\rm ho}})\propto N^{-2/5}$,
generally a very small quantity.
$R=a_{\rm ho}(15a_2/a_{\rm ho})^{1/5}N^{1/5}$ is the TF radius.
Typically, the dephasing rate is a fraction
of $\omega_{\perp}$ for a condensate of $\sim 10^5$ atoms
and $\lambda\sim 1$, i.e. its macroscopic spin direction is
lost in a few cycles of trap oscillation.
Figure \ref{fig3} also indicates that
the faster a ferromagnetic condensate diffuses its pointing
direction along the precession direction,
the larger and more tightly confined a condensate is.

When $B=0$, zero mode dynamics becomes much simpler
if the condensate spin direction is taken as z-axis.
Although there still exist more than one Goldstone mode
in this case, the mean field ground state corresponds to all
atoms in state $|+\rangle$ \cite{ho,ueda}. Thus to
leading order, the only relevant fluctuation is that
of the total number of atoms (corresponding to the U(1) symmetry),
and the phase diffusion dynamics becomes the same as in
a scalar condensate \cite{lewenstein}. Our theory as
developed in this paper does reduce to this simple limit.

In conclusion, we have studied in detail quantum phase
diffusions of a spinor-1 condensate with ferromagnetic
interactions. The condensate
ground state is very simple, all three spin components
have the same spatial mode function and their
associated phase functions are also identical.
We have constructed the zero mode Hamiltonian for the
condensate number and phase fluctuations,
and solved for the ground state distribution
of these fluctuations when both $N$ and ${\cal M}$ are conserved.
Furthermore, we have obtained analytically the
dynamic number and phase fluctuations
relating to both the quantum phase diffusion and the initial
distribution of these fluctuations.
We have identified a quantum phase diffusion coefficient
for the pointing direction of the condensate spin
and recovered its small-time quadratic t-dependent spreading.

This work is supported by the NSF grant No. PHYS-0140073
and by a grant from NSA, ARDA, and DARPA
under ARO Contract No. DAAD19-01-1-0667.

\appendix
\section{Coefficients in Eq. (12)}\label{coeff}
The coefficients in the zero mode Hamiltonian (\ref{hzero}) are
\begin{eqnarray}
a&=&\frac{1}{4n_0^2}\Big[\frac{\varepsilon}{3}(5+3m^2)
+2{\cal O}_{\phi\theta}\big[3c_0(1-m^2)^2\nonumber\\
&&\;\;\;\;\;\;\;\;\;+\frac{8}{3}c_2(1-3m^2)\big]\Big],\nonumber\\
b&=&\frac{1}{2n_0^2}\left[\varepsilon(1+m^2)
-8c_2{\cal O}_{\phi\theta}m^2\right]\nonumber\\
c&=&\frac{1}{3n_0^2}(\varepsilon-4c_2{\cal O}_{\phi\theta}),\nonumber\\
\alpha&=&\frac{\sqrt{2}}{6n_0^2}(1+3m^2)
(\varepsilon-4c_2{\cal O}_{\phi\theta}),\nonumber\\
\beta&=&-\frac{2m}{\sqrt{3}n_0^2}(\varepsilon
-4c_2{\cal O}_{\phi\theta}),\nonumber\\
\gamma&=&-\frac{2m}{n_0^2}\sqrt{\frac{2}{3}}\left[
\varepsilon-c_2{\cal O}_{\phi\theta}(1+3m^2)\right],\nonumber\\
\eta&=&-3c_2n_0^2{\cal O}_{\phi\phi}/\hbar^2.\nonumber
\end{eqnarray}

\end{document}